\documentclass[12pt, letterpaper]{article}

\usepackage[T1]{fontenc}
\usepackage{amsmath,amssymb}  
\usepackage{amsthm}
\usepackage{bm}      
\usepackage{dsfont}  
\usepackage{float}
\usepackage[pdftex]{graphicx}
\usepackage{geometry}
\usepackage[hypcap=true]{caption}
\usepackage{xcolor}
\definecolor{link}{rgb}{0.25,0.0,0.75}
\usepackage[%
   colorlinks=true,citecolor=blue,urlcolor=blue,linkcolor=link%
   ]{hyperref}

\usepackage{natbib}
   \bibpunct{[}{]}{,}{a}{}{,}

\usepackage[parfill]{parskip}    
\begin{document}

\thispagestyle{empty}
\begin{center}

\vspace*{0.5in}
{\LARGE
Adjusted closing prices
}

\vspace*{0.5in}
Vic Norton\\
Department of Mathematics and Statistics\\
Bowling Green State University\\
\url{mailto:vic@norton.name}\\
\url{http://vic.norton.name}

\vspace*{0.25in}
21--May--2010

\vspace*{1.0in}
ABSTRACT\\[2.5ex]
\begin{minipage}{4in}
Historical returns depend on historical closing prices and distributions. We describe
how to compute adjusted closing prices from closing price/distribution data with an
emphasis on spreadsheet implementation. Then the growth of a security from one
date to another (1 + \mbox{total return}) is just the ratio of the corresponding
adjusted closing prices.

\end{minipage}

\end{center}

\newpage
\setcounter{page}{1}
\section{The Growth Ratio}

If $C_-$ and $C_+$ are the closing prices of a given security on two successive business days, 
then the value of the security grows by $\sigma_+ - 1$ (decimal, percent) over the second
business day, where $ \sigma_+ = C_+/C_- $. Here we assume that the second business day
is not an ex-day corresponding to a declared distribution or split.

If day two is an ex-dividend day for a distribution of $D_+$ dollars per share, then the security
starts from an effective price of $C_- - D_+$ at the end of the first day
and grows to the $C_+$ price over the course of day two.
In this case the  \emph{growth ratio} $\sigma_+$ should be defined as
\begin{equation*}\label{sigma_d}
  \sigma_+ = \frac{C_+}{C_- - D_+}\tag{1d}
\end{equation*}
with the percentage growth again being $\sigma_+ - 1$.
Of course equation \eqref{sigma_d} can apply to a non-ex-day as well. Just set $D_+$ equal to 
zero.
After an $\alpha$\,:\,1 split the effective starting price is $C_-/\alpha$, and the
growth ratio is
\begin{equation*}
  \sigma_+ = \alpha\times\frac{C_+}{C_-}~. \tag{1s}
\end{equation*}
Splits can be also thought of as distributions. An $\alpha$\,:\,1 split
corresponds to a ~$D_+ = \frac{\alpha-1}{\alpha}\times C_-$ distribution.

\section{Adjusted Closing Prices}

Start with sequences of successive closing prices $C_i ~(i = 0,1,\ldots,n)$
and distributions $D_i ~(i = 1,\ldots,n)$ of a
given security. Define the corresponding growth ratios by
\begin{equation}\label{sigma}
  \sigma_i = \frac{C_i}{C_{i-1} - D_i}\quad(i = 1,\ldots,n).
\end{equation}
We will refer to any sequence of positive numbers $x_i ~(i = 0,1,\ldots,n)$
as \emph{adjusted closing prices} for the security if
\begin{equation}\label{xprices}
   \frac{x_i}{x_{i-1}} = \sigma_i\quad\mbox{for}\quad i = 1,\ldots,n~.
\end{equation}
Then the \emph{growth ratio} of the security for the duration under consideration is
\begin{equation}\label{growthratio}
  \sigma = \sigma_1\times\cdots\times\sigma_n = \frac{x_n}{x_0}~,
\end{equation}
and the percentage growth or total return is $\sigma - 1$.

See \citet{Investopedia:rm} for a less abstract description of how to compute the
adjusted closing price of a stock.

\section{Dividend Reinvestment}

If one buys $s_0$ shares of a security at closing price $C_0$, then the value of
the investment is $s_0C_0$. Suppose that dividends are reinvested automatically
and that the number of shares and the closing prices over the next $n$ business days are
$s_i$ and $C_i ~(i=1,\dots,n)$, respectively. Then the growth ratio $\sigma_i$ of the
investment over the $i$-th business day should clearly be the ratio of the
value of the investment at the close of day $i$ to its value at the close of day $i-1$:
\begin{equation}\label{sigma_reinv}
  \sigma_i = \frac{s_iC_i}{s_{i-1}C_{i-1}}~.
\end{equation}
By equating the $\sigma_i$ of \eqref{sigma} with the $\sigma_i$ of \eqref{sigma_reinv}
we see that
\begin{equation}\label{deltasi}
   s_i = s_{i-1} + \Delta s_i~,\quad \Delta s_i = \frac{D_i}{C_{i-1}-D_i}~.
\end{equation}
Thus the new shares $\Delta s_i$ are precisely what can be purchased with the dividend
$D_i$ at the ``ex-closing-price'' $C_{i-1}-D_i$.

\section{Spreadsheet Considerations}

Suppose successive closing prices $C$, distributions $D$, and
growth ratios $\sigma$ (as computed by \eqref{sigma}) are recorded as
three columns of a spreadsheet. The formulas
\begin{align}
  x_+ &= x_-\times\sigma_+ \label{fill+}\tag{FILL+}\\
  x_- &= x_+\div\sigma_+\label{fill-}\tag{FILL\hspace*{0.15ex}--}
\end{align}
then allow one to fill in adjusted closing prices from an arbitrary base adjusted
closing price on any given business day. If the $+=$ time direction is
up, enter the right-hand formula of the \eqref{fill+} equation in the cell
above the base price (now $x_-$) and ``Fill Up'' from there, and enter the
right-hand formula of the \eqref{fill-} equation in the cell below the base price
(now $x_+$) and ``Fill Down'' from this cell. In this way all adjusted closing
prices can be in filled in, up or down, from the base adjusted closing price.
(If the $+=$ time direction is down, one should ``Fill Down'' with \ref{fill+}
and ``Fill Up'' with \ref{fill-}.)

The following spreadsheet image illustrates the procedure. The adjusted
closing price of iShares Barclays 1-3Yr Tearsury Bond Fund (\texttt{SHY}) is
arbitrarily set at 100.000 on December 14, 2007.
All other adjusted closing prices are filled up or down from this base price
using \eqref{fill+} or \eqref{fill-}, respectively.
(The empty distribution cells are computed as zero.)

\begin{figure}[H]
\caption{Filling a spreadsheet\hspace*{6ex}}
  \label{spreadsheet}
  \centering
  \hspace{4.0ex}\includegraphics[scale=1.0]{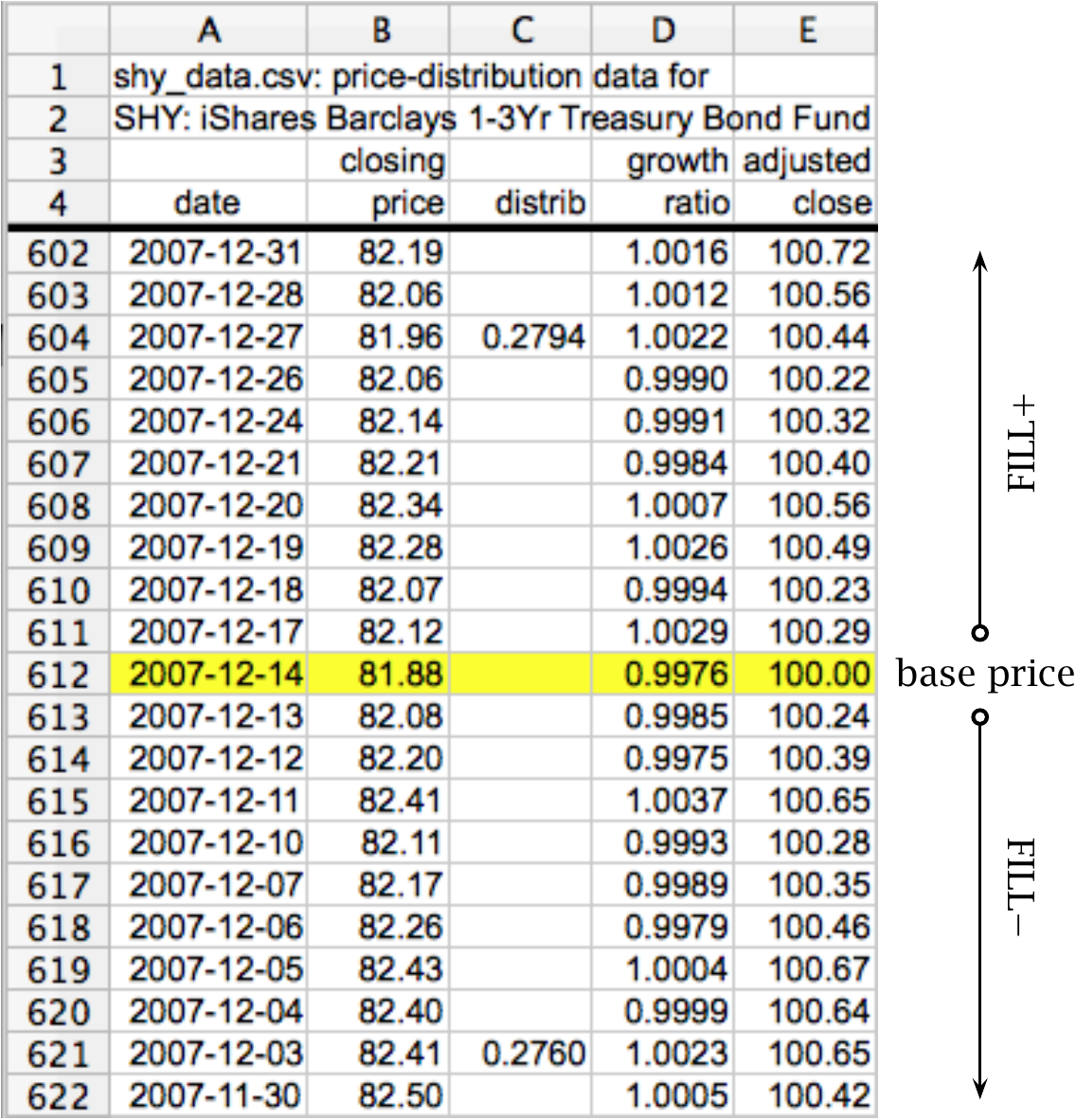}
\end{figure}

The growth of \texttt{SHY} in December 2007 is measured by the ratio of
its 2007-12-31 adjusted closing price to its 2007-11-30 price,
\[\frac{100.72}{100.42}=1.003.\]
Thus \texttt{SHY} gained 0.3\% in December 2007. The ratio, 1.003,
is also the product
of the December growth ratios: $1.0023 \,\times\cdots\times\, 1.0016$.

\section{Data}
\begin{figure}[H]
\caption{Adjusted closing prices\\
with base price $= 100$ on 2006-12-29}
  \label{adjclose_graph}
  \centering
  \includegraphics{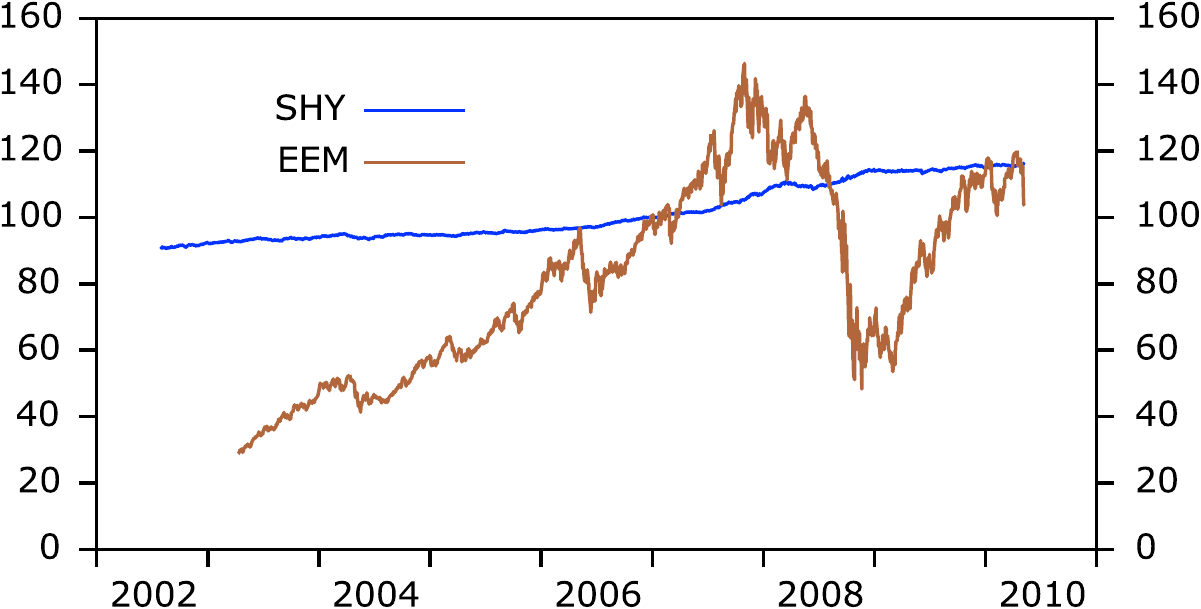}
\end{figure}

Here are two comma-separated-value (spreadsheet) files,
\href{http://vic.norton.name/finance-math/adjclose/shy\_data.csv}{shy\_data.csv}
and
\href{http://vic.norton.name/finance-math/adjclose/eem\_data.csv}{eem\_data.csv},
with closing price, distribution, and adjusted closing price data for
\begin{tabbing}
  \hspace*{5ex}\= \kill
  \>\=\texttt{SHY:}\=~iShares Barclays 1-3Yr Treasury Bond Fund \\
  \>\=\texttt{EEM:}\=~iShares MSCI Emerging Markets Index Fund
\end{tabbing}
respectively. The graphs of these adjusted closing price data are shown in Figure
\ref{adjclose_graph}.

We have chosen the bond fund \texttt{SHY} because has lots distributions to work with, generally one per month. On the other hand the equity fund \texttt{EEM} illustrates how
splits are handled; the shares of \texttt{EEM} split 3 for 1 in 2005 and again in 2008.

One can get historical closing prices for any security at
\href{http://finance.yahoo.com/}{Yahoo!\,FINANCE}
or more specifically, for our funds, at \\[1.0ex]
\hspace*{3ex}\url{http://finance.yahoo.com/q/hp?s=SHY+Historical+Prices} \\[0.5ex]
\hspace*{3ex}\url{http://finance.yahoo.com/q/hp?s=EEM+Historical+Prices} \\[1.0ex]
Dividends and splits can be found here as well, but unfortunately some dividends
may be missing. For example Yahoo!\,FINANCE
is missing the 27-Dec-2007 \texttt{SHY}
dividend of \$0.2794 per share shown in Figure \ref{spreadsheet}.

Yahoo!\,FINANCE also gives adjusted closing prices, but,
because of missing dividends, these prices may be inaccurate. Consider \texttt{SHY}
for example. The Yahoo's 2007-12-31 adjusted closing price is less than its 2007-11-30
adjusted closing price. Thus Yahoo!\,FINANCE would have you believe that \texttt{SHY}
lost money in December 2007, when, in fact, \texttt{SHY}
gained 0.3\% in this month---as we have just seen.

The best place to get iShares distribution data is from the horse's mouth: go to\\[1.0ex]
\hspace*{3ex}\url{http://us.ishares.com/product_info/fund/}\\[1.0ex]
click on the desired fund and then on the Distributions tab. In our case\\[1.0ex]
\hspace*{3ex}\url{http://us.ishares.com/product_info/fund/distributions/SHY.htm} \\[0.5ex]
\hspace*{3ex}\url{http://us.ishares.com/product_info/fund/distributions/EEM.htm} \\[1.0ex]
will get you all distributions of \texttt{SHY} and \texttt{EEM}, respectively.

\section{Closing Remarks}

All of this material probably appears elsewhere. We just don't know where.

\end{document}